\documentclass[english,british]{elsarticle}
\usepackage[T1]{fontenc}
\usepackage[latin9]{inputenc}
\usepackage[letterpaper]{geometry}
\usepackage{array}
\usepackage{float}
\usepackage{booktabs}
\usepackage{units}
\usepackage{amsthm}
\usepackage{amsmath}
\usepackage{graphicx}
\usepackage{setspace}
\usepackage{esint}
\makeatletter

\providecommand{\tabularnewline}{\\}

\newtheorem{defn}{Definition}
\usepackage{subfig} 
\biboptions{sort&compress}

\@ifundefined{showcaptionsetup}{}{%
 \PassOptionsToPackage{caption=false}{subfig}}
\usepackage{subfig}
\makeatother

\usepackage{babel}

\begin{document}
\selectlanguage{english}%

\title{A mathematical framework for reducing the domain in the mechanical
analysis of periodic structures}

\author{N. V. De Carvalho$^{\text{1}}$, S. T. Pinho$^{\text{1}}$, P. Robinson$^{\text{1}}$}
\address{$^{\text{1}}$Dept. of Aeronautics, Imperial College London, South
Kensington, London, SW7 2AZ, UK}

\begin{abstract}
A theoretical framework is developped leading to a sound derivation
of Periodic Boundary Conditions (PBCs) for the analysis of domains
smaller then the Unit Cells (UCs), named reduced Unit Cells (rUCs),
by exploiting non-orthogonal translations and symmetries. A particular
type of UCs, Offset-reduced Unit Cells (OrUCs) are highlighted. These
enable the reduction of the analysis domain of the traditionally defined
UCs without any loading restriction. The relevance of the framework
and its application to any periodic structure is illustrated through
two practical examples: 3D woven and honeycomb. 
\end{abstract}
\maketitle

\section{Introduction}

\begin{flushleft}
Numerical analysis of periodic materials and structures has proven
to be an extremely powerful tool. It provides detailed information,
such as failure initiation sites and stress-strain at smaller scales
(meso/micro) It has been successfully used to determine homogenised
properties, study the detailed stress-strain fields at nano- and microscopic
scales to obtain structural damage initiation conditions and sites,
as well as to simulate damage development and associated deterioration
of the homogenised mechanical properties \citep{Lomov2007}. Several
works can be found discussing the application of periodic boundary
conditions to representative regions, e.g. \citep{Xia2003,Xia2006,Sun1996,Li2008}.
For periodic structures, the Unit Cell (UC) is used as the representative
region, and the analysis is performed by applying periodic displacement
boundary conditions. The topological complexity of many UCs found
in practice, such as in typical woven composites, often leads to unpractical
modelling and analysis times. For this reason, internal symmetries
of the UCs must whenever possible be exploited to reduce the analysis
domain further (provided the appropriate boundary conditions are applied),
thus reducing both modelling and analysis time.
\par\end{flushleft}

\begin{flushleft}
A comprehensive study on the determination of reduced Unit Cells (rUCs)
for UD and particle reinforced composites was performed by \citet{Li1999,Li2001}
and \citet{Li2004}. Different rUCs, loading cases and correspondent
boundary conditions were determined and presented in detail. Applied
to textile composites, \citet{XiaodongTang07012003} proposed a general
framework for determining rUCs.
\par\end{flushleft}

\begin{flushleft}
In the first part of the present work, the derivation of the framework
proposed in \citep{XiaodongTang07012003} \foreignlanguage{british}{is
}revisited and some of its building blocks redefined, resulting in
a different, formally defined and more concise formulation. The framework
proposed by \citet{XiaodongTang07012003} requires the distinction
of two different cases of equivalence between subcells: (i) equivalence
is obtained by a symmetry operation or a translation, and (ii) equivalence
is obtained by the combination of a symmetry operation and a translation.
In the second case an additional vector of constants $\mathbf{r}$
(see \citet{XiaodongTang07012003}) needs to be considered when applying
the boundary conditions. The non-zero components of this vector are
tabulated for different cases and are determined by the FEM as part
of the solution. The formulation derived in the present work is more
generic, in that no cases need to be treated separately, and mathematically
complete in that no vector $\mathbf{r}$ needs to be determined from
tabulated data. All terms in the equation that assigns the periodic
boundary conditions for a rUC are fully defined, simplifying the formulation
and consequently their use.
\par\end{flushleft}

In the second part of this paper, the application of the formulation
developed and its potential is illustrated through two practical examples:
3D woven composites and honeycombs. Additionally, particular attention
is given to Offset-reduced Unit Cells as they allow the domain reduction
without load restrictions.

\section{Equivalence framework}

\begin{flushleft}
In this section, the equivalence framework is formally defined. It
is based on four concepts: physical equivalence, load equivalence,
periodicity and load admissibility. In the following sections each
of these concepts is detailed.
\par\end{flushleft}

\subsection{Physical equivalence}

\begin{flushleft}
Consider a domain $\mathbf{D}$ in space and within it a sub-domain
$E$. The latter has a defined boundary, Local Coordinate System (LCS)
$O_{E}xyz$, and a certain spatial distribution of $n$ physical properties
$\mathbf{P}^{i}$ with $i\in\left\{ 1,...,n\right\} $ . Each of these
physical properties $\mathbf{P}^{i}$ are expressed as a tensor written
in the LCS of $E$, i.e. $\mathbf{P}_{E}^{i}$. 
\par\end{flushleft}

\begin{defn} 

\begin{flushleft}
Two distinct sub-domains $E$ and $\hat{E}$ are \textit{Physically
Equivalent}, denoted:\textit{\begin{equation}
E\widehat{=}\hat{E}\label{eq:PhysicalEq}\end{equation}
} if for every point $A$ in $E$ there is a point $\hat{A}$ in $\hat{E}$
such that, for each physical property $i$, \begin{equation}
\mathbf{x}_{E}^{A}=\mathbf{x}_{\hat{E}}^{\hat{A}}\,\wedge\,\mathbf{P}_{E}^{i}\left(A\right)=\,\mathbf{P}_{\hat{E}}^{i}\left(\hat{A}\right)\label{eq:PhysicalCond}\end{equation}
 and vice-versa.
\par\end{flushleft}

\end{defn} 

\begin{flushleft}
In Eq. \ref{eq:PhysicalCond}, $\mathbf{x}_{E}^{A}$ and $\mathbf{x}_{\hat{E}}^{\hat{A}}$
are the coordinate vectors of $A$ and $\hat{A}$ given in the LCS
$O_{E}xyz$ and $O_{\hat{E}}xyz$ associated with $E$ and $\hat{E},$
respectively, Fig. \ref{fig:GeomRel}. The points $A$ and $\hat{A}$
for which Eq. \ref{eq:PhysicalEq} is verified are designated as physically
equivalent points. %
\begin{figure}[H]
\selectlanguage{british}%
\begin{centering}
\includegraphics[scale=0.5]{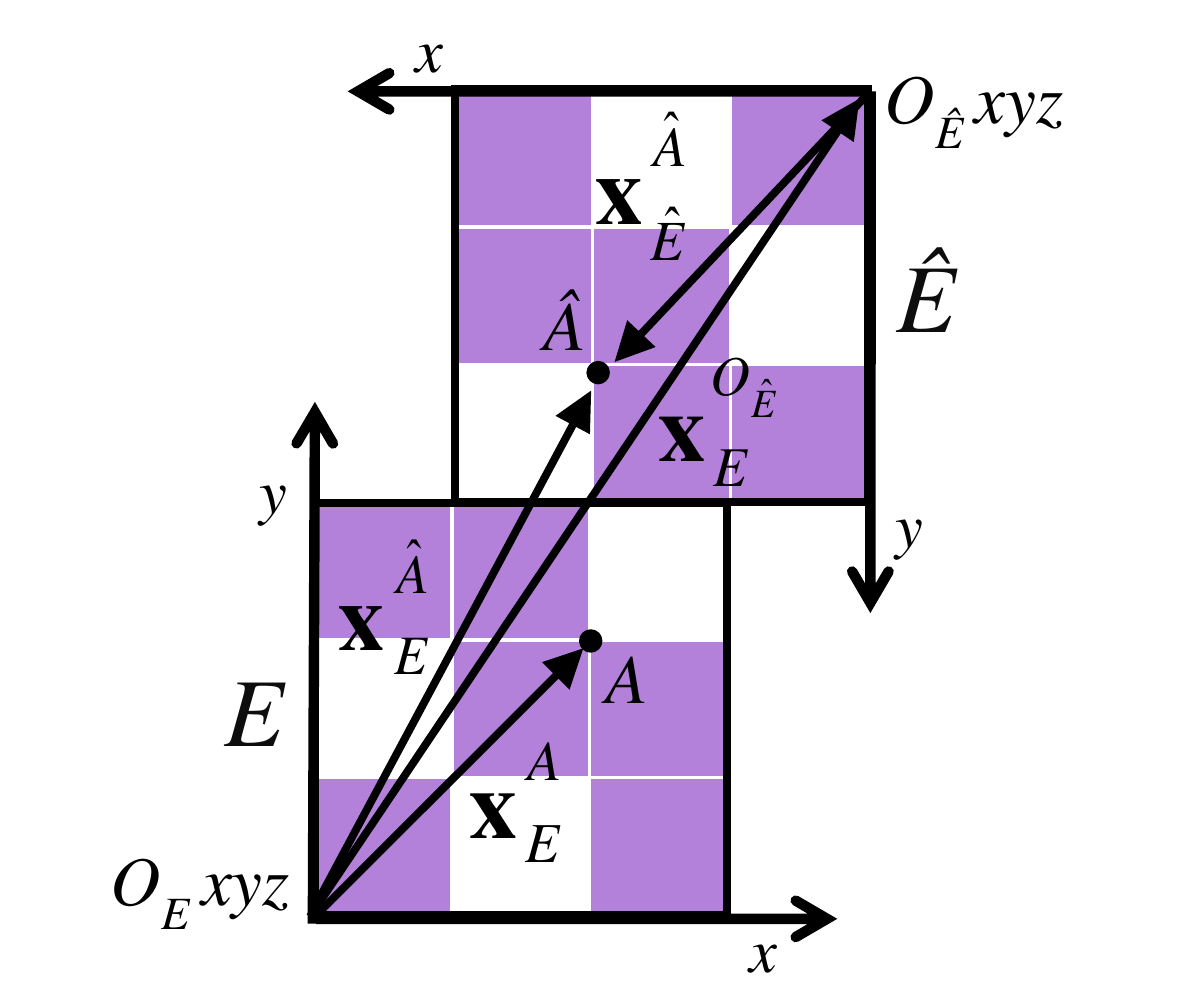}
\par\end{centering}

\selectlanguage{english}%
\caption{\selectlanguage{british}%
\label{fig:GeomRel}Geometrical relation between equivalent points. \selectlanguage{english}
}

\end{figure}

\par\end{flushleft}

\subsection{Periodicity and Unit Cell \label{sub:Per_Def}}

\begin{flushleft}
Across the literature, different definitions can be found for periodic
structure and UC. In the present work, periodic structure and UC are
defined based on the concept of physical equivalence.
\par\end{flushleft}

\begin{defn} A domain $D$ is periodic if it can be reconstructed
by tessellation of, non-overlapping, physically equivalent sub-domains
$E_{i}$ with parallel LCS, i.e. if for all $i\neq j$:\begin{equation}
E_{i}\widehat{=}E_{j}\wedge O_{E_{i}}xyz\parallel O_{E_{j}}xyz\label{eq:Periodicitydef2}\end{equation}
 The smallest sub-domain verifying the periodicity definition is designated
as an Unit Cell. 

\end{defn}

\subsection{Loading equivalence}

\begin{flushleft}
The concept of load equivalence (see \citet{XiaodongTang07012003})
provides a relation between physically-equivalent sub-domains, once
the structure they are part of is loaded. Let us consider a periodic
structure as defined in the previous section. 
\par\end{flushleft}

\begin{defn} 

\begin{flushleft}
Load equivalence between two physically equivalent points $A$ and
$\hat{A}$ is verified if the strains and stresses at these points,
given in the LCS of the sub-domains, can be related by:\begin{equation}
\varepsilon_{E}\left(A\right)=\gamma\varepsilon_{\hat{E}}\left(\hat{A}\right)\label{eq:LoadEqstrain}\end{equation}
 \begin{equation}
\sigma_{E}\left(A\right)=\gamma\sigma_{\hat{E}}\left(\hat{A}\right),\label{eq:LoadEqstress}\end{equation}
 where the load reversal factor, $\gamma=\pm1$, is used to enforce
the equivalence between fields of physically equivalent sub-domains. 
\par\end{flushleft}

\end{defn} 

\begin{flushleft}
For Eqs. \ref{eq:LoadEqstrain} and \ref{eq:LoadEqstress} to hold,
the length scale of the loading variation must be larger than the
length scale of the sub-domains, such that an approximately periodic
variation of the strains and stress fields is assured. If a structure
is entirely composed by load equivalent sub-domains, its response
can be obtained by analysing one of these domains alone, instead of
analysing the entire structure. However, the appropriate boundary
conditions have to be applied. These guarantee that the sub-domain,
although isolated, has the same response it would have if it was embedded
in the structure. 
\par\end{flushleft}

\subsection{Sub-domain admissibility}

\begin{flushleft}
Not all physically equivalent sub-domains can be used to analyse the
response of a periodic structure under all loading conditions. The
use of sub-domains smaller than the UC to analyse the response of
a periodic structure is restricted by the relations between the LCS
of these sub-domains. The sufficient and necessary condition for admissibility
of a sub-domain to be used in the analysis of a periodic structure
is derived below. 
\par\end{flushleft}

\subsubsection*{Average and fluctuation fields}

\begin{flushleft}
For convenience, the strain field of a sub-domain at a point $A$
is decomposed as the sum of a volume average and a fluctuation term,
see Fig.\ref{fig:Fluct_Avg_1D}:\begin{align}
\varepsilon\left(A\right) & =\left\langle \varepsilon\right\rangle +\varepsilon^{*}\left(A\right),\label{eq:Average_Fluct_Strain}\end{align}
 where $\left\langle \bullet\right\rangle =\frac{1}{V}\int_{V}\bullet dV$
is the volume average operator over the volume $V$, and $\varepsilon^{*}$
is the fluctuation term, see \citet{Suquet1987}. It is possible to
find the displacements at a given point by integration of Eq. \ref{eq:Average_Fluct_Strain}.
Assuming small displacements and no rigid body rotations, the displacement
relative to the origin of a LCS, attached to the subdomain, comes
as: \begin{equation}
\mathbf{u}\left(A\right)=\left\langle \varepsilon\right\rangle \cdot\mathbf{x}^{A}+\mathbf{u}^{*}\left(A\right).\label{eq:Average_Fluct_Disp}\end{equation}
\begin{figure}[H]
\selectlanguage{british}%
\begin{centering}
\includegraphics[scale=0.35]{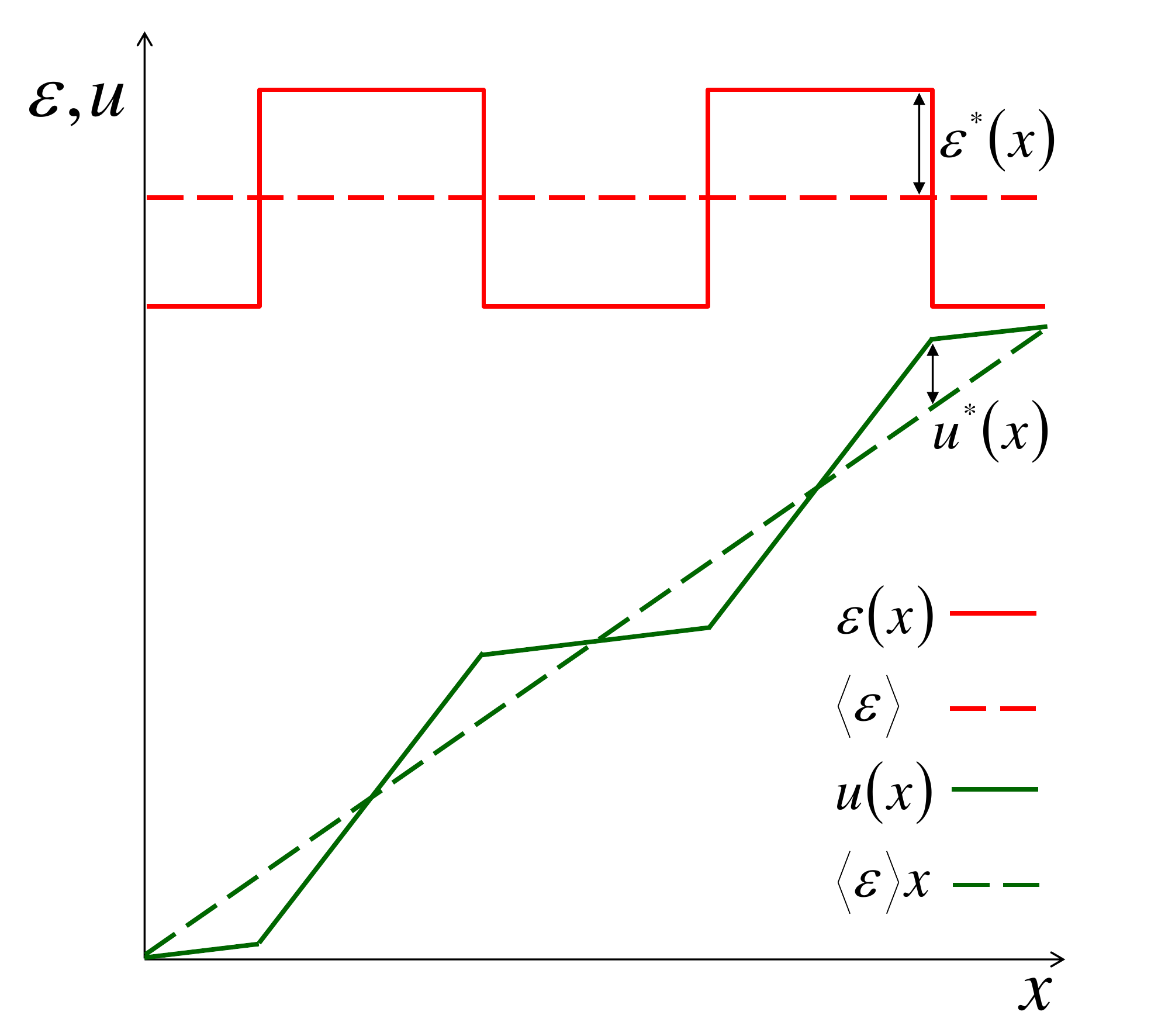}
\par\end{centering}

\selectlanguage{english}%
\caption{\label{fig:Fluct_Avg_1D}Idealised relation between the fluctuation
and average fields of strain and displacement in a periodic structure.}

\selectlanguage{british}%
\selectlanguage{english}

\end{figure}

\par\end{flushleft}

\subsubsection*{Relations between fields of two equivalent points in a global coordinate
system}

\begin{flushleft}
Knowing that the coordinate vectors of two equivalent points $A$
and $\hat{A}$ given in their LCSs are identical, Eq. \ref{eq:PhysicalCond},
they can be related in the LCS of the sub-domain $E$ by:\begin{equation}
\mathbf{x}_{E}^{A}=\mathbf{T}\left(\mathbf{x}_{E}^{\hat{A}}-\mathbf{x}_{E}^{O_{\hat{E}}}\right)\label{eq:PositionGlobalA_Ahat}\end{equation}
 where $\mathbf{T}$ is the transformation matrix between the LCSs
of $\hat{E}$ and $E$, and $\mathbf{x}_{E}^{O_{\hat{E}}}$ is the
position vector of the origin of the LCS of the sub-domain $\hat{E}$
given in the LCS of the sub-domain $E$, Fig. \ref{fig:GeomRel}.
Similarly, using Eq. \ref{eq:LoadEqstrain}, the strains at two equivalent
points can be related in the LCS of $E$ by:\begin{equation}
\varepsilon_{E}\left(A\right)=\gamma\mathbf{T}\varepsilon_{E}\left(\hat{A}\right)\mathbf{T}^{t}.\label{eq:StrainGlobalA_Ahat}\end{equation}

\par\end{flushleft}

\begin{flushleft}
The relation between the volume average of the strain of the equivalent
sub-domains $E$ and $\hat{E}$, in the LCS of the first, can be obtained
directly by integrating Eq. \ref{eq:StrainGlobalA_Ahat}:\begin{equation}
\left\langle \varepsilon\right\rangle _{E}^{E}=\gamma\mathbf{T}\left\langle \varepsilon\right\rangle _{E}^{\hat{E}}\mathbf{T}^{t},\label{eq:Strain_Avg_Global}\end{equation}

\par\end{flushleft}

\begin{flushleft}
where the lower index of $\left\langle \bullet\right\rangle $ refers
to the coordinate system, and the upper index to the domain over which
the volume average was taken. Decomposing the strain field in Eq.
\ref{eq:StrainGlobalA_Ahat} into its average and fluctuation parts
and using Eq. \ref{eq:Strain_Avg_Global}, the relation between the
strain fluctuations field of two equivalent points is obtained:\begin{equation}
\varepsilon_{E}^{*}\left(A\right)=\gamma\mathbf{T}\varepsilon_{E}^{*}\left(\hat{A}\right)\mathbf{T}^{t}.\label{eq:Strain_Fluct_Global}\end{equation}
In general, the displacement of a point $A$ can be obtained from:\begin{align}
\mathbf{u}(\mathbf{x}^{A}) & =\mathbf{u}_{0}+\int_{\mathbf{x}=0}^{\mathbf{x}^{A}}d\mathbf{u}=\mathbf{u}_{0}+\int_{\mathbf{x}=0}^{\mathbf{x}^{A}}\nabla.\mathbf{u}d\mathbf{x}=\mathbf{u}_{0}+\int_{\mathbf{x}=0}^{\mathbf{x}^{A}}\varepsilon d\mathbf{x}+\int_{\mathbf{x}=0}^{\mathbf{x}^{A}}\Omega d\mathbf{x,}\label{eq:DispAfromInt}\end{align}
 where $\varepsilon=\frac{1}{2}\left(\nabla.\mathbf{u}+\nabla.\mathbf{u}^{t}\right)$,
$\Omega=\frac{1}{2}\left(\nabla.\mathbf{u}-\nabla.\mathbf{u}^{t}\right)$
and $\mathbf{x}^{A}$ is the coordinate vector of point $A$. Considering
the structure has no rigid body motion $\mathbf{u}_{0}=0$ , nor rotation
$\Omega=0$, the displacement of a point $A$ can be simply obtained
by:\begin{equation}
\mathbf{u}(\mathbf{x}^{A})=\int_{\mathbf{x}=0}^{\mathbf{x}^{A}}\varepsilon d\mathbf{x}.\label{eq:DispAfromIntStrain}\end{equation}
 The strain fluctuations of two equivalent points belonging to the
sub-domains $E$ and $\hat{E}$ are related in the LCS of $E$ by
(see also Eq. \ref{eq:Strain_Fluct_Global} ):\begin{equation}
\varepsilon_{E}^{*}\left(\mathbf{x}_{E}^{A}\right)=\gamma\mathbf{T}\varepsilon_{E}^{*}\left(\mathbf{x}_{E}^{\hat{A}}\right)\mathbf{T}{}^{t}.\label{eq:Strain_fluct}\end{equation}
 Knowing that two equivalent points are related by (see also Eq. \ref{eq:PositionGlobalA_Ahat}):
\[
\mathbf{x}_{E}^{A}=\mathbf{T}\left(\mathbf{x}_{E}^{\hat{A}}-\mathbf{x}_{E}^{O_{\hat{E}}}\right),\]
 and that the displacement at the origin is equal to zero, integrating
Eq. \ref{eq:Strain_fluct} it is possible to obtain:\begin{align}
\int_{\mathbf{x}_{E}=0}^{\mathbf{x}_{E}^{A}}\varepsilon_{E}^{E*}\left(\mathbf{x}\right)d\mathbf{x} & =\int_{\mathbf{x}_{E}=0}^{\mathbf{x}_{E}^{A}}\gamma\mathbf{T}\varepsilon_{E}^{\hat{E}*}\left(\mathbf{T}^{t}\mathbf{x}_{E}+\mathbf{x}_{E}^{O_{\hat{E}}}\right)\mathbf{T}^{t}\mathbf{T}d\mathbf{x}_{E}\nonumber \\
\mathbf{u}_{E}^{*}(\mathbf{x}_{E}^{A}) & =\gamma\mathbf{T}\left\{ \mathbf{u}_{E}^{*}(\mathbf{T}^{t}\mathbf{x}_{E}^{A}+\mathbf{x}_{E}^{O_{\hat{E}}})-\mathbf{u}_{E}^{*}(\mathbf{x}_{E}^{O_{\hat{E}}})\right\} .\label{eq:Disp_fluct_1}\end{align}
 Equation \ref{eq:Disp_fluct_1} provides a relation between the displacement
perturbations of two equivalent points given in the LCS of one of
the sub-domains. Apart from $\mathbf{u}^{*}(\mathbf{x}_{E}^{O_{\hat{E}}})$,
all variables are known; below it is shown that $\mathbf{u}^{*}\left(\mathbf{x}_{E}^{O_{\hat{E}}}\right)=0$.
\par\end{flushleft}

According to the definition of periodicity, a periodic structure can
be reconstructed from physically equivalent sub-domains with parallel
coordinate systems. The strain fields at two equivalent points belonging
to different sub-domains are related by:\begin{equation}
\varepsilon_{E}\left(\mathbf{x}_{E}^{A}\right)=\gamma\mathbf{T}\varepsilon_{E}\left(\mathbf{T}^{t}\mathbf{x}_{E}^{A}+\mathbf{x}_{E}^{O_{\hat{E}}}\right)\mathbf{T}^{t}.\label{eq:StrainGlobalA_Ahat_Ap}\end{equation}
 If we consider that the sub-domains are UCs, since the coordinate
systems are parallel, the matrix $\mathbf{T}$ will be equal to the
identity matrix. Moreover, all equivalent sub-domains will be admissible
and have a load reversal factor $\gamma=1$. Equation \ref{eq:StrainGlobalA_Ahat_Ap}
can then be simply written as:\begin{equation}
\varepsilon_{E}\left(\mathbf{x}_{E}^{A}\right)=\varepsilon_{E}\left(\mathbf{x}_{E}^{A}+\mathbf{x}_{E}^{O_{\hat{E}}}\right).\label{eq:StrainGlobalA_Ahat_ApUC}\end{equation}
 If the two UCs being considered are adjacent, a vector $\mathbf{d}=\mathbf{x}_{E}^{O_{\hat{E}}}$
can be defined and Eq. \ref{eq:StrainGlobalA_Ahat_ApUC} can be generalized
for any point $\mathbf{x}$ of the structure:\begin{equation}
\varepsilon\left(\mathbf{x}\right)=\varepsilon\left(\mathbf{x}+\mathbf{d}\right),\label{eq:StrainGlobalGen_pUC}\end{equation}
 where $\mathbf{d}$ is commonly named the periodicity vector, and
corresponds to the period of the function $\varepsilon\left(\mathbf{x}\right)$
\citep{Suquet1987}.

The integral of a periodic function $f$ of period $D$ can always
be written as:\begin{equation}
\int f\left(t\right)dt=g\left(t\right)+\bar{f}t+C,\label{eq:periodicFprep}\end{equation}
 where $g\left(t\right)$ is also a periodic function of period $D$,
$\overline{f}$ is the average the periodic function $f$, and $C$
is a constant \citep{Elle2001}. Using Eq. \ref{eq:periodicFprep}
it is possible to write:\begin{equation}
\mathbf{u}^{*}\left(\mathbf{x}\right)=\int\varepsilon^{*}\left(\mathbf{x}\right)d\mathbf{x}=\mathbf{h}\left(\mathbf{x}\right).\label{eq:dispfluct_def}\end{equation}

The average term that would appear in Eq. \ref{eq:dispfluct_def}
is zero since by definition $\varepsilon^{*}\left(\mathbf{x}\right)$
has zero average. Additionally, knowing that at the origin $\mathbf{u}^{*}\left(\mathbf{x}\right)$
is equal to zero, one can conlude that $C$ will also be zero and
thus $\mathbf{u}^{*}\left(\mathbf{x}\right)$ will be a periodic function
with zero average. Using the above result, and knowing that the integration
over a period of a periodic function with zero average is equal to
zero, one can integrate both sides of Eq. \ref{eq:Disp_fluct_1} over
a period:\begin{align}
\int_{\mathbf{x}}^{\mathbf{x}+\mathbf{d}}\mathbf{u}^{*}(\mathbf{x})d\mathbf{x} & =\gamma\mathbf{\int_{\mathbf{x}}^{\mathbf{x}+\mathbf{d}}T}\left\{ \mathbf{u}^{*}(\mathbf{T}^{t}\mathbf{x}+\mathbf{x}_{E}^{O_{\hat{E}}})-\mathbf{u}^{*}(\mathbf{x}_{E}^{O_{\hat{E}}})\right\} d\mathbf{x}\nonumber \\
0 & =\gamma\int_{\mathbf{x}}^{\mathbf{x}+\mathbf{d}}\mathbf{u}^{*}(\mathbf{x}_{E}^{O_{\hat{E}}})d\mathbf{x},\label{eq:Disp_fluct_Or_Zero}\end{align}
 obtaining:\begin{equation}
\mathbf{u}^{*}(\mathbf{x}_{E}^{O_{\hat{E}}})=0.\label{eq:Disp_fluct_Or_zeroFinal}\end{equation}
 Substituting Eq. \ref{eq:Disp_fluct_Or_zeroFinal} in Eq. \ref{eq:Disp_fluct_1},
the relation between the displacement perturbations at two equivalent
points can be finally obtained:\begin{equation}
\mathbf{u}_{E}^{*}(\mathbf{x}_{E}^{A})=\mathbf{\gamma T}\mathbf{u}_{E}^{*}(\mathbf{x}_{E}^{\hat{A}}).\label{eq:Disp_Fluct_Global_Ap}\end{equation}

\subsubsection*{Evaluation of the sub-domain admissibility}

For a sub-domain to be admissible, the volume average (homogenised)
strain calculated for this sub-domain on a given reference system
must equal that volume average on any other sub-domain (on the same
reference system), as the volume average is a homogenised entity,
hence independent of the sub-domain where it was calculated. From
load equivalence, the strains at physically-equivalent points are
related (Eq. 5). Eq. 11 is obtained by simply integrating this relation
over the sub-domain, but does not enforce directly that the volume
average strain is a macroscopic entity independent of the particular
sub-domain. For the sub-domain to be admissible, the following condition
must be verified:\begin{equation}
\left\langle \varepsilon\right\rangle _{E}^{E}=\left\langle \varepsilon\right\rangle _{E}^{\hat{E}}\label{eq:VolAvgEq}\end{equation}

\begin{flushleft}
as, for a sub-domain to be admissible, the homogenised strain on a
given reference system (in this case $E$) must be the same for any
sub-domain (in this case $E$ and $\hat{E}$). Therefore, Eq. \ref{eq:Strain_Avg_Global}
with Eq. \ref{eq:VolAvgEq} lead to the condition of sub-domain admissibility,
as defined below. 
\par\end{flushleft}

\begin{defn}

\begin{flushleft}
A given sub-domain $E$ is admissible for the analysis of a periodic
structure under a given loading $\left\langle \varepsilon\right\rangle _{E}$,
if $\mathbf{T}_{i}$ and $\gamma_{i}$ correspondent to any other
sub-domain $\hat{E}_{i}$ are such that, for all $\hat{E}_{i}$:\begin{equation}
\left\langle \varepsilon\right\rangle _{E}^{E}=\gamma_{i}\mathbf{T}_{i}\left\langle \varepsilon\right\rangle _{E}^{\hat{E}_{i}}\mathbf{T}_{i}^{t}\label{eq:sub_domain_adm_gen}\end{equation}

\par\end{flushleft}

\end{defn}

\begin{flushleft}
Equation \ref{eq:sub_domain_adm_gen} can be used to, for a given
loading, determine the load reversal factors $\gamma_{i}$ associated
with each of the sub-domains. The admissibility of a subdomain for
structural analysis leads to the definition of a rUC.
\par\end{flushleft}

\begin{defn}

\begin{flushleft}
A reduced Unit Cell is a domain, smaller than the Unit Cell, that
can be used to determine the response of a periodic structure to a
given loading. The condition to be verified by a reduced Unit Cell
in structural analysis is defined by Eq. \ref{eq:sub_domain_adm_gen}.
\par\end{flushleft}

\end{defn}

\section{Derivation of Periodic Boundary conditions}

\begin{flushleft}
To ensure the response of a periodic structure under a given loading
can be determined from the response of a rUC, the appropriate boundary
conditions that must applied to the latter need to be determined.
In this section, the equivalence framework, presented previously,
is used to derive the periodic boundary conditions for the analysis
of rUCs. 
\par\end{flushleft}

\begin{flushleft}
Consider two adjacent sub-domains $E$ and $\hat{E}$ that are physically
and load equivalent. If a point $\hat{A}$ belonging to $\hat{E}$
is chosen to be at the boundary of the sub-domain $E$, then its equivalent
point $A$ is also be at the boundary of $E$. Since both points $A$
and $\hat{A}$ belong to $E$, the displacement at each point can
be obtained using Eq. \ref{eq:Average_Fluct_Disp}:\begin{eqnarray}
\mathbf{u}\left(A\right) & = & \left\langle \varepsilon\right\rangle \mathbf{x}^{A}+\mathbf{u}^{*}\left(A\right)\label{eq:DispA}\\
\mathbf{u}\left(\hat{A}\right) & = & \left\langle \varepsilon\right\rangle \mathbf{x}^{\hat{A}}+\mathbf{u}^{*}\left(\hat{A}\right)\label{eq:DispAhat}\end{eqnarray}
 All quantities in Eqs. \ref{eq:DispA} and \ref{eq:DispAhat} are
written in the LCS of $E$, and the volume average is taken over the
sub-domain $E$ (the subscript will be omitted hereafter for convenience).
Since both points are equivalent, their positions are related by Eq.
\ref{eq:PositionGlobalA_Ahat} leading to:\begin{equation}
\mathbf{u}\left(A\right)=\left\langle \varepsilon\right\rangle \mathbf{T}\left(\mathbf{x}^{\hat{A}}-\mathbf{x}^{O_{\hat{E}}}\right)+\mathbf{u}^{*}\left(A\right)\label{eq:DispConstrain1}\end{equation}
 Knowing that the displacement fluctuations at two equivalent points
are related by Eq. \ref{eq:Disp_Fluct_Global_Ap}, if Eq. \ref{eq:DispAhat}
is multiplied by $\gamma\mathbf{T}$ and then subtracted to Eq. \ref{eq:DispA},
the displacement fluctuations cancel, leading to:\begin{align}
\mathbf{u}\left(A\right)-\gamma\mathbf{T}\mathbf{u}\left(\hat{A}\right)=\left(\left\langle \varepsilon\right\rangle \mathbf{T}-\gamma\mathbf{T}\left\langle \varepsilon\right\rangle \right)\mathbf{x}^{\hat{A}}-\left\langle \varepsilon\right\rangle \mathbf{T}\mathbf{x}^{O_{\hat{E}}}\label{eq:DispConstrain2}\end{align}
 Provided the sub-domain $\hat{E}$ is admissible, see Definition
4, the term $\left(\left\langle \varepsilon\right\rangle \mathbf{T}-\gamma\mathbf{T}\left\langle \varepsilon\right\rangle \right)$
is zero. Using this result, Eq. \ref{eq:DispConstrain2} can be simplified
to Eq. \ref{eq:DispConstrainFinal}, which is the main outcome of
this analysis and can be used directly to apply periodic boundary
conditions to a sub-domain:\begin{eqnarray}
\mathbf{u}\left(A\right)-\gamma\mathbf{T}\mathbf{u}\left(\hat{A}\right) & = & -\left\langle \varepsilon\right\rangle \mathbf{Tx}^{O_{\hat{E}}}\label{eq:DispConstrainFinal}\end{eqnarray}
 Once a displacement constraint equation is associated to all points
at the boundary of the sub-domain $E$, loading can be applied by
prescribing a volume average strain $\left\langle \varepsilon\right\rangle $. 
\par\end{flushleft}

\begin{flushleft}
It is relevant to notice that the displacement constraint equation
traditionally used to impose periodic boundary conditions on a UC,
see \citet{Suquet1987} for example, is a particular case of Eq. \ref{eq:DispConstrainFinal}
where the matrix $\mathbf{T}$ is equal to the identity matrix $\mathbf{I}$,
since the LCSs of the UCs are parallel by definition and consequently,
from the sub-domain admissibility evaluation, the load reversal factor
is equal to one. It is important to highlight the differences between
the result obtained above and the one obtained in \citet{XiaodongTang07012003};
as referred before, Eq. \ref{eq:DispConstrainFinal} is completely
generic and self sufficient: no distinction needs to be made, in the
current formulation, between the type of operation needed to achieve
equivalence between subdomains. Moreover, all terms in Eq. \ref{eq:DispConstrainFinal}
are fully defined, and can therefore be readily used to prescribe
periodic boundary conditions to a given subdomain.
\par\end{flushleft}

\subsection{Offset reduced Unit Cells}

\begin{flushleft}
According to the periodicity definition given in \ref{sub:Per_Def},
a UC is the smallest sub-domain that allows a periodic structure to
be reconstructed by tessellation of sub-domains that are physically
equivalent to the UC and have parallel LCS. Nevertheless, in most
applications the UC is defined such that the LCS are not only parallel
but orthogonally translated, Fig. \ref{fig:Unit Cell}. However, smaller
UCs can in general be defined if non-orthogonal translations are considered,
Fig. \ref{fig:OrUC}. Although, according to the definition, the representative
sub-domains obtained through non-orthogonal translation are in fact
UCs, in the present paper they are referred to as Offset-reduced Unit
Cells, since they lead to a reduction in the domain of the traditionally
defined UCs, Fig. \ref{fig:Unit Cell vs OrUC}.
\par\end{flushleft}

\begin{flushleft}
An important feature of OrUCs is that all loading combinations are
admissible. This key feature has been identified by \citet{Li2001}
and used in the derivation of PBCs for rUCs of UD composites, and
cracked laminates \citep{Li2009}. It is relevant to highlight that
using the present formulation this feature comes as a natural result:
since the LCS of all sub-domains are parallel, they relate to each
other by the identity matrix, i.e. $\mathbf{T}=\mathbf{I}$, as a
consequence Eq. \ref{eq:sub_domain_adm_gen} is always verified and
therefore all loading cases are admissible.%
\begin{figure}[H]
\hfill{}\subfloat[\selectlanguage{british}%
\label{fig:Unit Cell}Unit Cell\selectlanguage{english}
]{\includegraphics[scale=0.4]{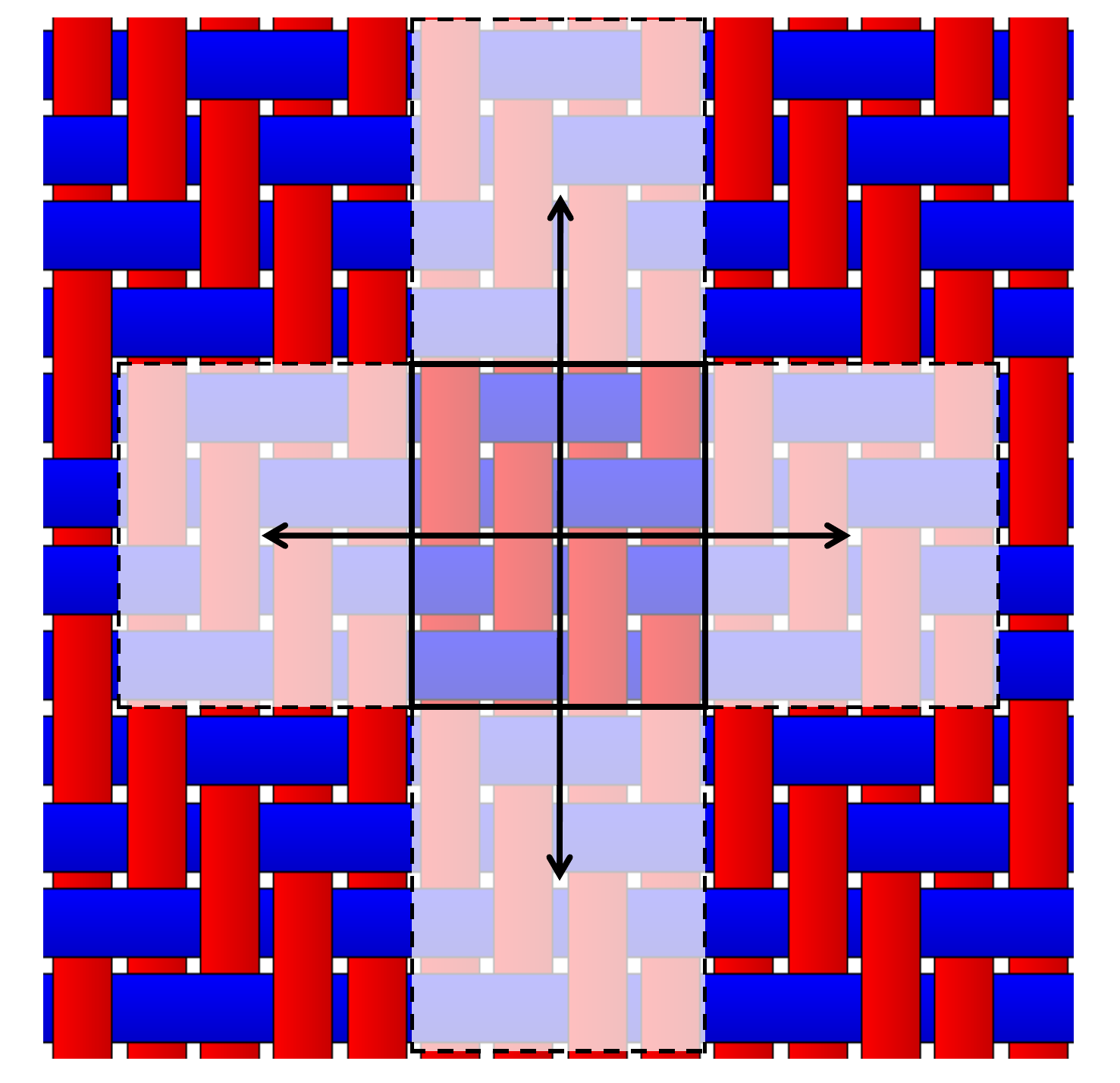}

\selectlanguage{british}%
\selectlanguage{english}
}\hfill{}\subfloat[\selectlanguage{british}%
\label{fig:OrUC}\foreignlanguage{english}{Offset reduced Unit Cell}\selectlanguage{english}
]{\includegraphics[scale=0.4]{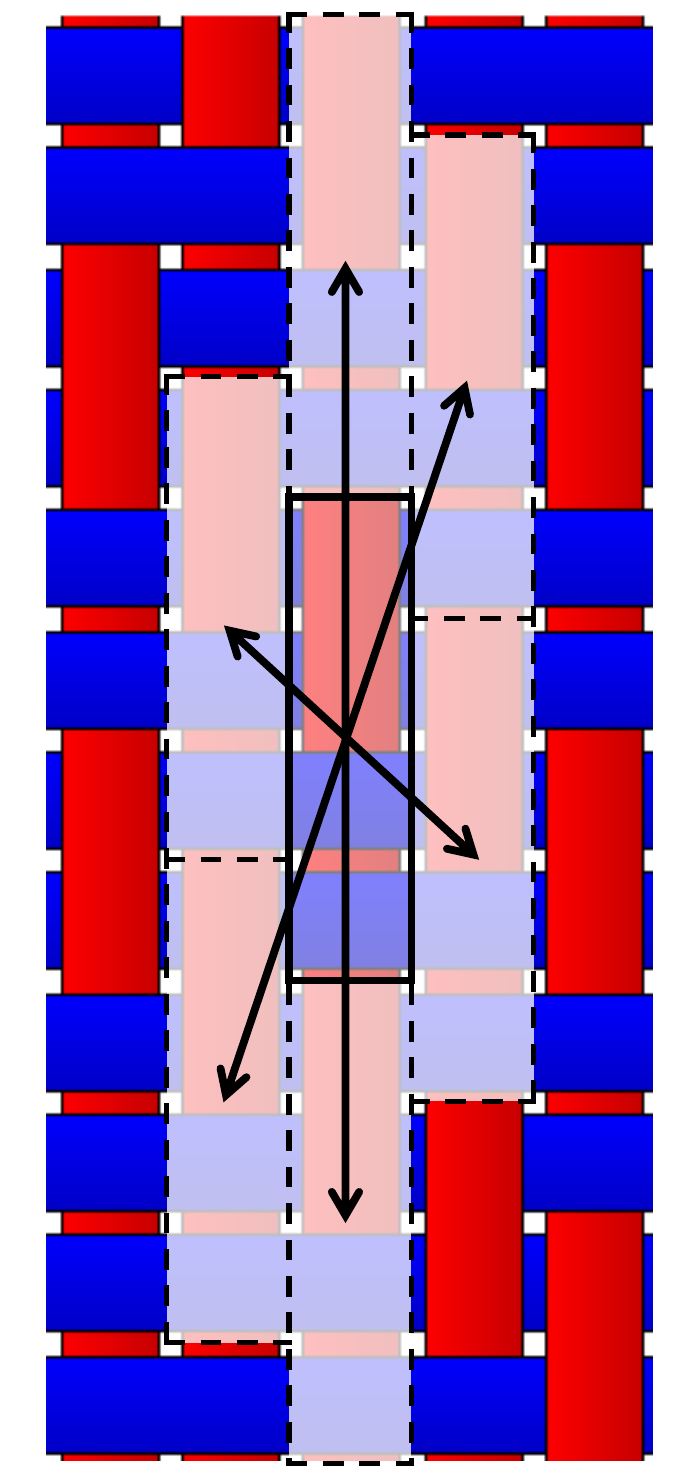}

\selectlanguage{british}%
\selectlanguage{english}
}\hfill{}\caption{\selectlanguage{british}%
\label{fig:Unit Cell vs OrUC}a) Unit Cell (UC) and b) Offset reduced
Unit Cell (OrUC) of a $2\times2$ Twill weave\selectlanguage{english}
}

\end{figure}

\par\end{flushleft}

\section{Applications}

In the present section two applications of the formulation presented
previously are illustrated.

\subsection{3D Woven Composites}

\begin{figure}[H]
\begin{centering}
\includegraphics[scale=0.4]{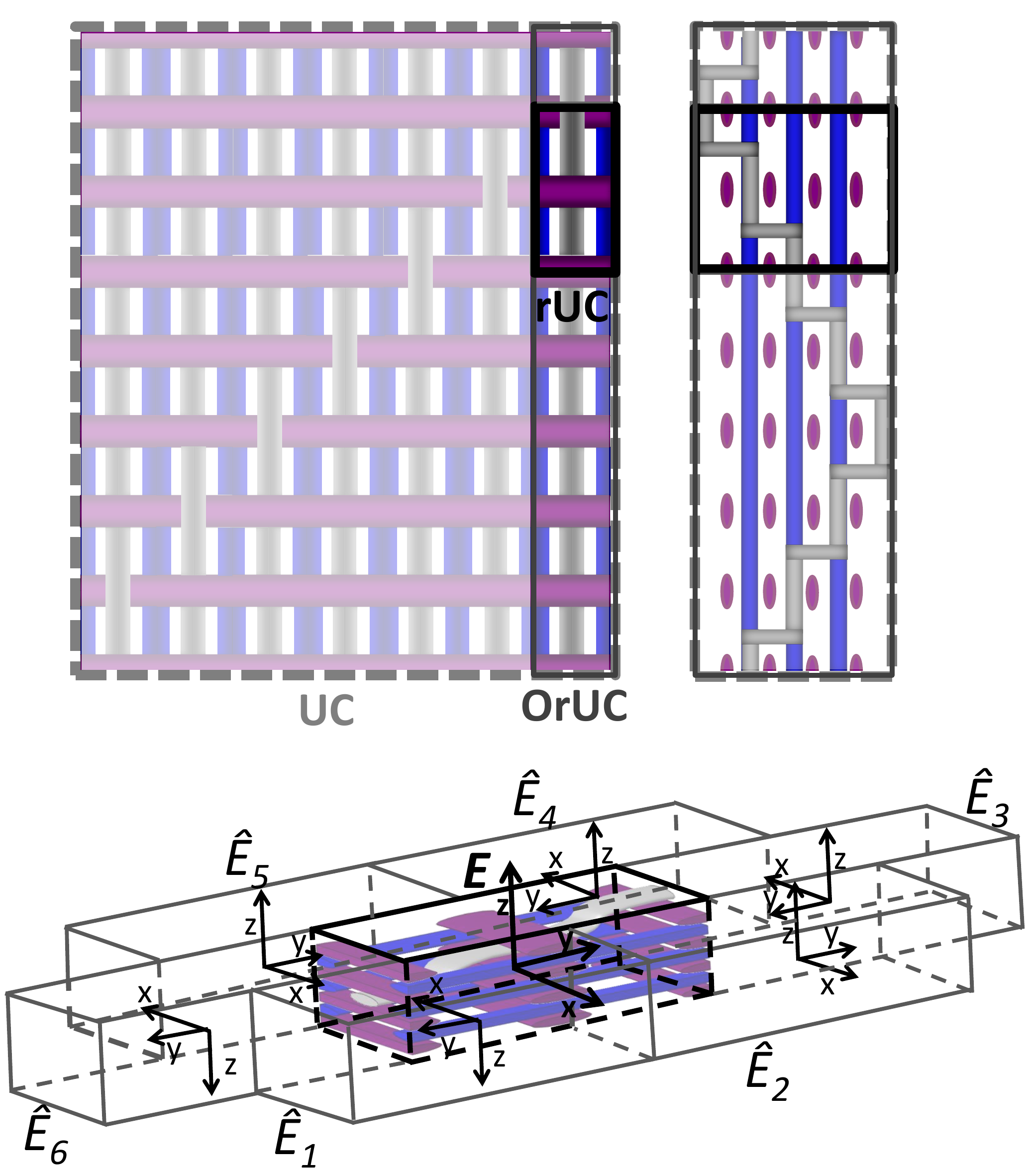} 
\par\end{centering}

\caption{\label{fig:rUC3D3D}UC, OrUC and rUC of a 3D woven reinforcement architecture;
representation of the reduced Unit Cell (rUC) and adjacent sub-domains. }

\end{figure}

\begin{flushleft}
The UCs of 3D woven composites can be significantly larger than their
2D counterparts, mostly due to the more intricate reinforcement architecture
and 3D nature. Therefore, the domain reduction enabled by the use
of rUCs can be very significant. Figure \ref{fig:rUC3D3D}, shows
an UC, an OrUC and a rUC of a given 3D woven architecture, highlighting
the domain reduction achieved: OrUC and rUC reduce the analysis domain
to $\nicefrac{1}{7}$ and $\nicefrac{1}{28}$ of the UC, respectively.
\par\end{flushleft}

\begin{figure}[H]
\begin{centering}
\includegraphics[scale=0.45]{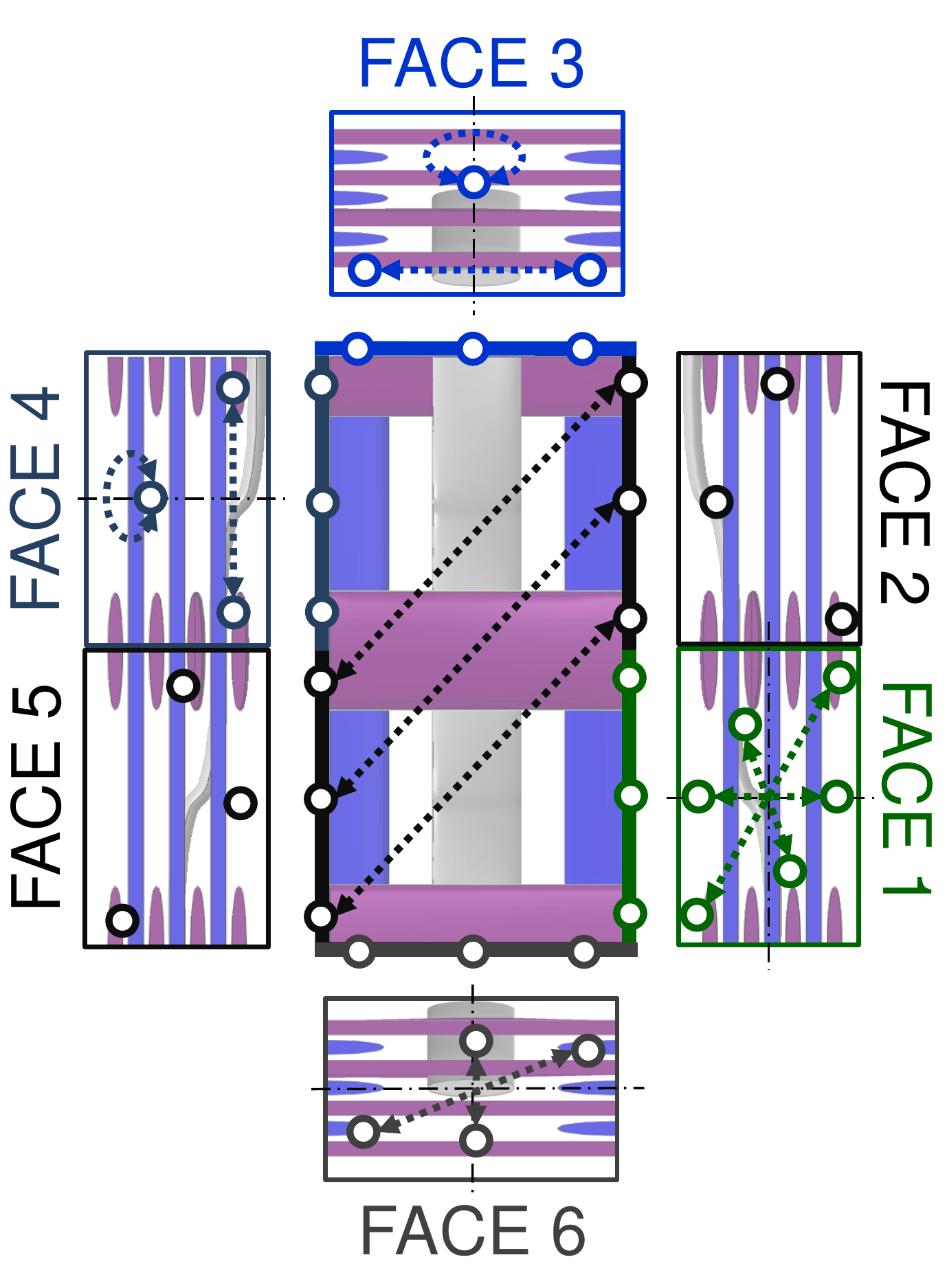}
\par\end{centering}

\caption{\label{fig:rUC3D_geom_rel}Geometrical relations between equivalent
points at the boundary of the rUC.}

\selectlanguage{british}%
\selectlanguage{english}

\end{figure}

To define the periodic boundary conditions for the analysis of the
rUC of Fig. \ref{fig:rUC3D3D}, Eq. \ref{eq:DispConstrainFinal},
the geometric relations between equivalent points at the rUC boundary
need to be found. These are obtained by applying Eq. \ref{eq:PositionGlobalA_Ahat}
to the equivalent domains at the boundary of the rUC and are given
in Table \ref{tab:rUC3Dgeom} and illustrated in Fig. \ref{fig:rUC3D_geom_rel}.
Since in general, $\mathbf{T\neq I}$, the load admissibility needs
to be evaluated and $\gamma$, for each adjacent subdomain, determined.
This is performed evaluating Eq. \ref{eq:sub_domain_adm_gen}, and
summarized in Table \ref{tab:rUC3D&AdmissibleLoadings}. Given a certain
loading and using the data from Tables \ref{tab:rUC3Dgeom} and \ref{tab:rUC3D&AdmissibleLoadings},
Eq. \ref{eq:DispConstrainFinal} can be fully defined and periodic
boundary conditions prescribed to the rUC. %
\begin{table}[H]
\caption{\label{tab:rUC3Dgeom}Geometrical relations between equivalent points
at the boundary for the 3D woven rUC. $l$, $w$ and $t$ are respectively,
the length width and and thickness of the rUC.}

\centering{}\begin{tabular}{>{\centering}p{0.55in}cccccc}
\toprule 
\selectlanguage{british}%
\selectlanguage{english}
 & {\scriptsize $\hat{E}_{1}$} & {\scriptsize $\hat{E}_{2}$} & {\scriptsize $\hat{E}_{3}$} & {\scriptsize $\hat{E}_{4}$} & {\scriptsize $\hat{E}_{5}$} & {\scriptsize $\hat{E}_{6}$}\tabularnewline
\midrule
\midrule 
{\scriptsize $\mathbf{T}$} & {\tiny $\left[\begin{array}{ccc}
-1 & 0 & 0\\
0 & -1 & 0\\
0 & 0 & -1\end{array}\right]$} & {\tiny $\left[\begin{array}{ccc}
1 & 0 & 0\\
0 & 1 & 0\\
0 & 0 & 1\end{array}\right]$} & {\tiny $\left[\begin{array}{ccc}
-1 & 0 & 0\\
0 & -1 & 0\\
0 & 0 & 1\end{array}\right]$} & {\tiny $\left[\begin{array}{ccc}
-1 & 0 & 0\\
0 & -1 & 0\\
0 & 0 & 1\end{array}\right]$} & {\tiny $\left[\begin{array}{ccc}
1 & 0 & 0\\
0 & 1 & 0\\
0 & 0 & 1\end{array}\right]$} & {\tiny $\left[\begin{array}{ccc}
-1 & 0 & 0\\
0 & -1 & 0\\
0 & 0 & -1\end{array}\right]$}\tabularnewline
\midrule 
{\scriptsize $\mathbf{x}^{O_{\hat{E}}}$} & {\tiny $\left[\begin{array}{c}
w\\
-\frac{l}{2}\\
0\end{array}\right]$} & {\tiny $\left[\begin{array}{c}
w\\
\frac{l}{2}\\
0\end{array}\right]$} & {\tiny $\left[\begin{array}{c}
0\\
l\\
0\end{array}\right]$} & {\tiny $\left[\begin{array}{c}
-w\\
\frac{l}{2}\\
0\end{array}\right]$ } & {\tiny $\left[\begin{array}{c}
-w\\
-\frac{l}{2}\\
0\end{array}\right]$} & {\tiny $\left[\begin{array}{c}
0\\
-l\\
0\end{array}\right]$}\tabularnewline
\midrule 
{\scriptsize $\mathbf{x}^{\hat{A}}$} & {\tiny $\left[\begin{array}{c}
x=\frac{w}{2}\\
-\frac{l}{2}\leq y\leq0\\
-\frac{t}{2}\leq z\leq\frac{t}{2}\end{array}\right]$} & {\tiny $\left[\begin{array}{c}
x=\frac{w}{2}\\
0\leq y\leq\frac{l}{2}\\
-\frac{t}{2}\leq z\leq\frac{t}{2}\end{array}\right]$} & {\tiny $\left[\begin{array}{c}
-\frac{w}{2}\leq x\leq\frac{w}{2}\\
y=\frac{l}{2}\\
-\frac{t}{2}\leq z\leq\frac{t}{2}\end{array}\right]$} & {\tiny $\left[\begin{array}{c}
x=-\frac{w}{2}\\
0\leq y\leq\frac{l}{2}\\
-\frac{t}{2}\leq z\leq\frac{t}{2}\end{array}\right]$} & {\tiny $\left[\begin{array}{c}
x=-\frac{w}{2}\\
-\frac{l}{2}\leq y\leq0\\
-\frac{t}{2}\leq z\leq\frac{t}{2}\end{array}\right]$} & {\tiny $\left[\begin{array}{c}
-\frac{w}{2}\leq x\leq\frac{w}{2}\\
y=-\frac{l}{2}\\
-\frac{t}{2}\leq z\leq\frac{t}{2}\end{array}\right]$}\tabularnewline
\midrule 
{\scriptsize $\mathbf{x}^{A}$} & {\tiny $\left[\begin{array}{c}
w-x_{1}^{\hat{A}}\\
-\frac{l}{2}-x_{2}^{\hat{A}}\\
-x_{3}^{\hat{A}}\end{array}\right]$} & {\tiny $\left[\begin{array}{c}
x_{1}^{\hat{A}}-w\\
x_{2}^{\hat{A}}-\frac{l}{2}\\
x_{3}^{\hat{A}}\end{array}\right]$} & {\tiny $\left[\begin{array}{c}
-x_{1}^{\hat{A}}\\
-x_{2}^{\hat{A}}+l\\
x_{3}^{\hat{A}}\end{array}\right]$} & {\tiny $\left[\begin{array}{c}
-x_{1}^{\hat{A}}-w\\
-x_{2}^{\hat{A}}+\frac{l}{2}\\
x_{3}^{\hat{A}}\end{array}\right]$} & {\tiny $\left[\begin{array}{c}
x_{1}^{\hat{A}}+w\\
x_{2}^{\hat{A}}+\frac{l}{2}\\
x_{3}^{\hat{A}}\end{array}\right]$} & {\tiny $\left[\begin{array}{c}
-x_{1}^{\hat{A}}\\
-x_{2}^{\hat{A}}-l\\
-x_{3}^{\hat{A}}\end{array}\right]$}\tabularnewline
\bottomrule
\end{tabular}
\end{table}
\begin{table}[H]
\caption{\label{tab:rUC3D&AdmissibleLoadings}Admissible loading cases and
respective value of the load reversal factor $\gamma_{i}$, correspondent
to each adjacent sub-domain $\hat{E}_{i}$, for the 3D woven rUC.}

\centering{}\begin{tabular}{>{\centering}p{1in}c>{\centering}p{1in}}
\toprule 
\selectlanguage{british}%
\selectlanguage{english}
 & {\scriptsize $\gamma_{i}$} & {\scriptsize Admissible loading}\tabularnewline
\midrule
\midrule 
{\scriptsize Case 1}  & {\scriptsize $\left[\begin{array}{cccccc}
1 & 1 & 1 & 1 & 1 & 1\end{array}\right]$} & {\scriptsize $\left[\begin{array}{ccc}
\sigma_{11} & \sigma_{12} & 0\\
\sigma_{21} & \sigma_{22} & 0\\
0 & 0 & \sigma_{33}\end{array}\right]$}\tabularnewline
\midrule 
{\scriptsize Case 2} & {\scriptsize $\left[\begin{array}{cccccc}
1 & 1 & -1 & -1 & 1 & 1\end{array}\right]$} & {\scriptsize $\left[\begin{array}{ccc}
0 & 0 & \sigma_{13}\\
0 & 0 & \sigma_{23}\\
\sigma_{31} & \sigma_{32} & 0\end{array}\right]$}\tabularnewline
\bottomrule
\end{tabular}
\end{table}

\subsection{Honeycombs}

\begin{flushleft}
Honeycombs are other example of an extensively used periodic structure
for which UC modelling and analysis can be simplified by the use of
rUCs. Figure \ref{fig:honeyUC&rUC} shows the UC and rUC for a honeycomb
structure. Following the procedure described previously, the geometrical
relations between equivalent points at the boundary are first determined,
Table \ref{tab:rUCHoney&geom}, and Figs \ref{fig:rUChoney_adj_sub}
and \ref{fig:rUChoney_geom_rel}. The load admissibility is evaluated
and the load reversal factors $\gamma_{i}$ found, Table \ref{tab:rUCHoney&AdmissibleLoad}.%
\begin{figure}[H]
\hfill{}\includegraphics[scale=0.4]{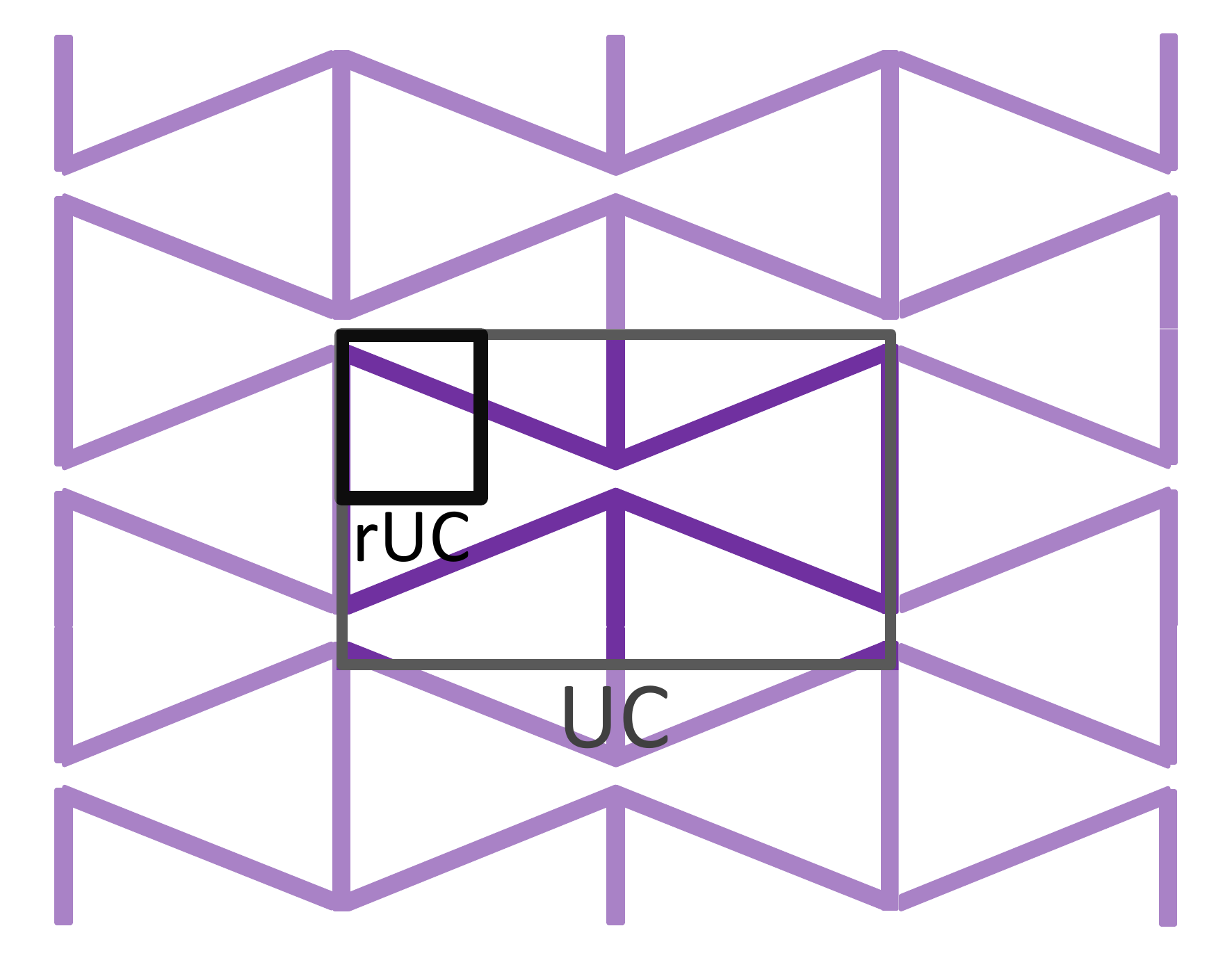}\hfill{}

\caption{\label{fig:honeyUC&rUC}Unit cell (UC) and reduced Unit Cell (rUC)
of a honeycomb structure.}

\end{figure}
\begin{figure}[H]
\hfill{}\subfloat[\label{fig:rUChoney_geom_rel}\foreignlanguage{british}{rUC and adjacent
sub-domains}]{\begin{centering}
\includegraphics[scale=0.35]{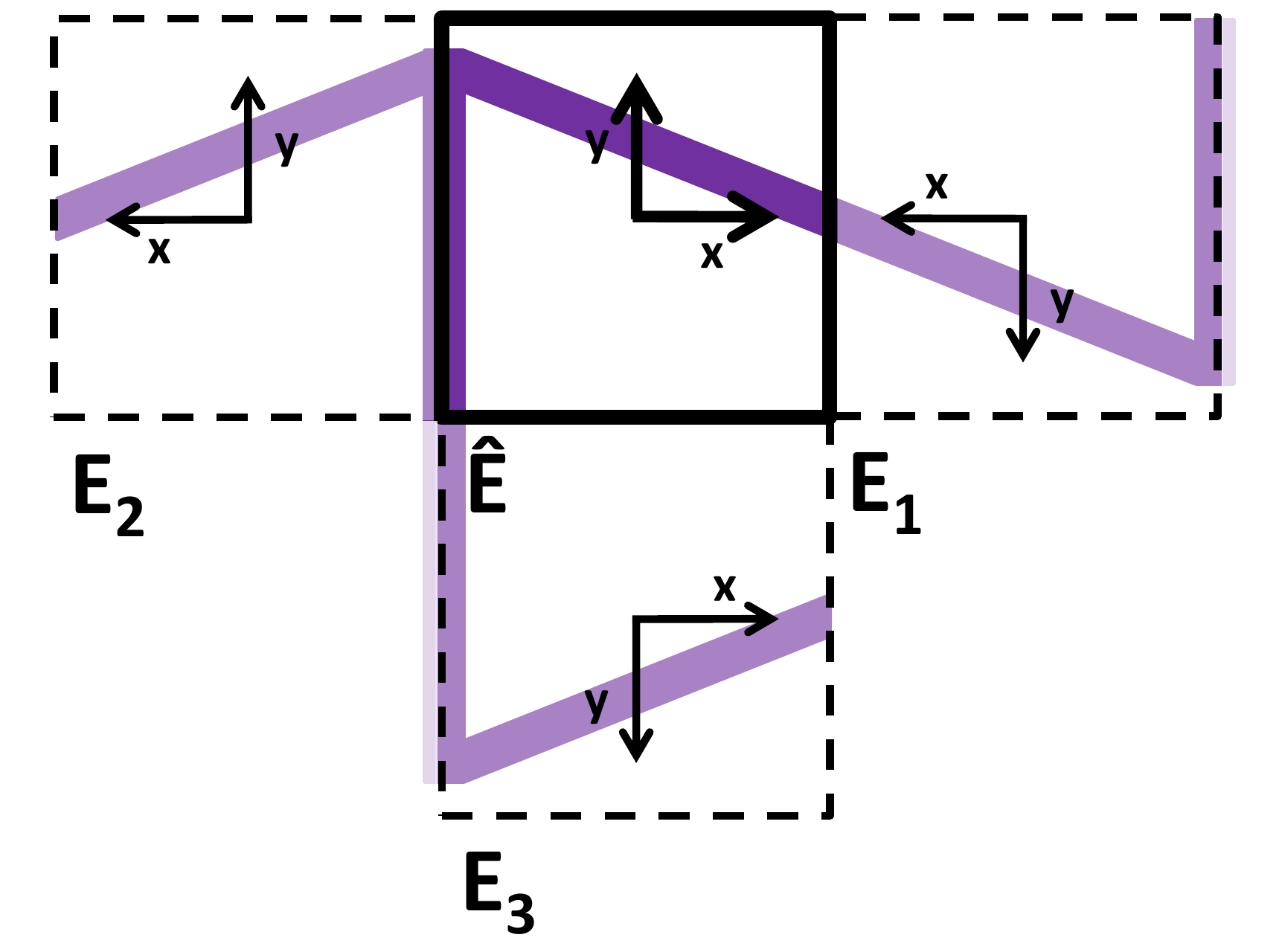}
\par\end{centering}

}\hfill{}\subfloat[\selectlanguage{british}%
Geometrical relations between equivalent points at the rUC boundary\selectlanguage{english}
]{\includegraphics[scale=0.3]{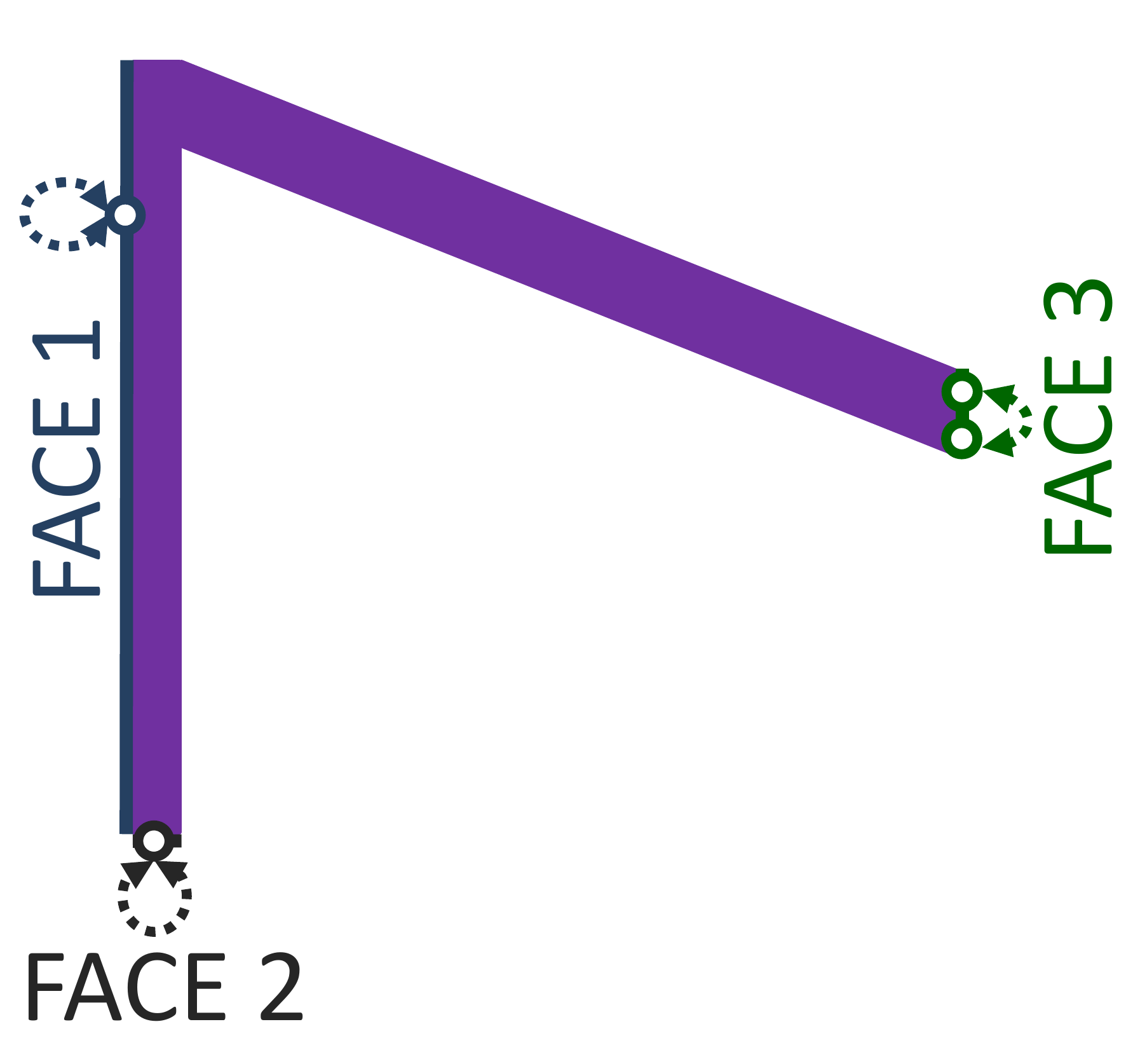}

}\hfill{}

\caption{\selectlanguage{british}%
\selectlanguage{english}
}

\selectlanguage{british}%
\selectlanguage{english}

\end{figure}
\begin{table}[H]
\caption{\label{tab:rUCHoney&geom}Geometrical relations between equivalent
points at the boundary of the honeycomb rUC. $l$, $w$ are respectively,
the length and width of the rUC.}

\centering{}\begin{tabular}{cccc}
\toprule 
\selectlanguage{british}%
\selectlanguage{english}
 & {\scriptsize $\hat{E}_{1}$} & {\scriptsize $\hat{E}_{2}$} & {\scriptsize $\hat{E}_{3}$}\tabularnewline
\midrule
\midrule 
{\scriptsize $\mathbf{T}$} & {\scriptsize $\left[\begin{array}{cc}
-1 & 0\\
0 & -1\end{array}\right]$} & {\scriptsize $\left[\begin{array}{cc}
-1 & 0\\
0 & 1\end{array}\right]$} & {\scriptsize $\left[\begin{array}{cc}
1 & 0\\
0 & -1\end{array}\right]$}\tabularnewline
\midrule 
{\scriptsize $x^{O_{\hat{E}}}$} & {\scriptsize $\left[\begin{array}{c}
w\\
0\end{array}\right]$} & {\scriptsize $\left[\begin{array}{c}
-w\\
0\end{array}\right]$} & {\scriptsize $\left[\begin{array}{c}
0\\
-l\end{array}\right]$}\tabularnewline
\midrule 
{\scriptsize $\mathbf{x}^{\hat{A}}$} & {\scriptsize $\left[\begin{array}{c}
x=\frac{w}{2}\\
-\frac{l}{2}\leq y\leq\frac{l}{2}\end{array}\right]$} & {\scriptsize $\left[\begin{array}{c}
x=-\frac{w}{2}\\
-\frac{l}{2}\leq y\leq\frac{l}{2}\end{array}\right]$} & {\scriptsize $\left[\begin{array}{c}
-\frac{w}{2}\leq x\leq\frac{w}{2}\\
y=-\frac{l}{2}\end{array}\right]$}\tabularnewline
\midrule 
{\scriptsize $\mathbf{x}^{A}$} & {\scriptsize $\left[\begin{array}{c}
-x_{1}^{\hat{A}}+w\\
-x_{2}^{\hat{A}}\end{array}\right]$} & {\scriptsize $\left[\begin{array}{c}
-x_{1}^{\hat{A}}-w\\
x_{2}^{\hat{A}}\end{array}\right]$} & {\scriptsize $\left[\begin{array}{c}
x_{1}^{\hat{A}}\\
-x_{2}^{\hat{A}}-l\end{array}\right]$}\tabularnewline
\bottomrule
\end{tabular}
\end{table}
\begin{table}[H]
\caption{\label{tab:rUCHoney&AdmissibleLoad}Admissible loading cases and respective
value of the load reversal factor $\gamma_{i}$, correspondent to
each adjacent sub-domain $\hat{E}_{i}$, for the rUC of the honeycomb.}

\centering{}\begin{tabular}{ccc}
\toprule 
\selectlanguage{british}%
\selectlanguage{english}
 & {\scriptsize $\gamma_{i}$} & {\scriptsize Admissible loading}\tabularnewline
\midrule
\midrule 
{\scriptsize Case 1} & {\scriptsize $\left[\begin{array}{ccc}
1 & 1 & 1\end{array}\right]$} & {\scriptsize $\left[\begin{array}{cc}
\sigma_{11} & 0\\
0 & \sigma_{22}\end{array}\right]$}\tabularnewline
\midrule 
{\scriptsize Case 2} & {\scriptsize $\left[\begin{array}{ccc}
1 & -1 & -1\end{array}\right]$} & {\scriptsize $\left[\begin{array}{cc}
0 & \sigma_{12}\\
\sigma_{21} & 0\end{array}\right]$}\tabularnewline
\bottomrule
\end{tabular}
\end{table}

\par\end{flushleft}

\section{Conclusions}

\begin{flushleft}
A theoretical framework leading to a sound derivation of PBCs for
the analysis of domains smaller then the Unit Cells (UCs), named reduced
Unit Cells (rUCs), by exploiting non-orthogonal translations and symmetries
whithin the UC was developped. The investment in defining the problem
formally resulted in a simple and readily usable formulation. The
method is applied to two different periodic structures illustrating
the potential of the rUC concept. Offset reduced Unit Cells are highlighted
as a particular case with interesting features, allowing the analysis
of domains smaller than the UC without any load restrictions.
\par\end{flushleft}

\begin{flushleft}
 \bibliographystyle{plain}
 
\par\end{flushleft}\selectlanguage{british}

\end{document}